# Machine Learning Assisted Design of Complex and High Entropy Alloys by Hybrid HiPIMS/Pulsed-DC PVD Process for Low Carbon Energy Applications in Extreme Environments

Paul Foulquier[1]*, Ryma Haddad[2], Ali Mahmoud[3], Eric Monsifrot[4], Fanny Balbaud-Célérier[2], Jean-Philippe Poli[3], Frédéric Schuster[5]

[1]*Université Paris-Saclay, CEA, INSTN, 91191 Gif-sur-Yvette, France*

[2]*Université Paris-Saclay, CEA, S2CM, 91191 Gif-sur-Yvette, France*

[3]*Université Paris-Saclay, CEA, LIST, 91191 Gif-sur-Yvette, France*

[4]*AZ-Concept SASU, 61 rue du Grand Faubourg, 25460 Etupes, France*

[5]*Université Paris-Saclay, CEA, Cross-cutting program – Materials and Processes, 91191 Gif-sur-Yvette, France*

**Corresponding author: paul.foulquier@cea.fr*

**Keywords** : Coating; HiPIMS; Artificial Intelligence; Combinatorial synthesis; High entropy materials; Harsh environments; Nuclear

## Abstract

Complex and high entropy alloys are attracting much attention currently thanks to their mechanical and corrosion resistance properties in harsh environments, in particular needed for carbon-free energy applications. However, their elaboration in bulk and in thin film form in a trial-and-error approach is impractical due to their complexity and the cocktail effect. The recent development of artificial intelligence brings a new possibility for their elaboration and adjustment of their properties.

Firstly, we present an overview of Materials and data science research. Then we describe how DIADEM - French initiative for Materials and Data science convergence – tackles the development of innovative coatings for carbon-free energy applications (nuclear, high temperature electrolysis, …) thanks to the development of a nationwide network of synthesis and characterization platforms – the DIADEM discovery hub. We describe in particular DIADEM-2D, an AI-driven Hybrid HiPIMS/Pulsed-DC PVD process using 4 cathodes in confocal combinatorial configuration. We present the chemical-system selection for high entropy alloy coatings using literature and experimental data on corrosion resistance in molten salt media and nuclear accidental conditions. A descriptor-based, element-agnostic machine learning framework linking deposition parameters and coating composition has been implemented, enabling prediction of the process parameters required to reach a target composition. We demonstrate the feasibility and accuracy of this composition-prediction and deposition-optimization strategy, and we discuss its current scope and limitations.

## Introduction

Low-carbon energy systems frequently operate under extreme environments where structural materials and coatings are simultaneously exposed to multiple degradation mechanisms. Nuclear energy is a particularly demanding example, as materials must withstand the combined effects of high-temperature corrosion or oxidation, severe mechanical loading, and intense neutron irradiation. These coupled phenomena often involve antagonistic material requirements, making the identification of optimal materials a genuine multi-objective optimization problem in which the best solution results from a compromise between several competing performance criteria rather than from the maximization of a single property.

Complex Concentrated Alloys (CCAs) and Refractory High-Entropy Alloys (RHEAs) have emerged as highly promising candidates for these applications owing to their exceptional compositional flexibility and their remarkable resistance to combined thermomechanical, chemical, and irradiation-induced degradation. However, the immense compositional space accessible to these alloys, together with the complexity of coating synthesis processes, makes conventional trial-and-error approaches impractical. Artificial Intelligence (AI), coupled with high-throughput experimentations, offers a powerful framework for accelerating materials discovery by identifying optimal trade-offs between multiple functional properties while drastically reducing the experimental effort.

In this article, we present how **DIADEM**—the French national initiative dedicated to the convergence of Materials Science and Data Science—addresses these challenges through the development of innovative protective coatings for carbon-free energy systems. The DIADEM Discovery Hub relies on a nationwide network of advanced synthesis, characterization, and data infrastructures designed to accelerate materials discovery.

Particular emphasis is placed on **DIADEM-2D**, an AI-driven hybrid HiPIMS/Pulsed-DC magnetron sputtering platform equipped with four independently controlled cathodes in a confocal combinatorial configuration. This unique experimental setup enables the rapid exploration of multidimensional compositional spaces and the generation of large, high-quality datasets for machine learning. A descriptor-based, element-agnostic predictive model linking deposition parameters, process conditions and coating composition has been developed to support inverse process design. This model addresses the prediction of deposition conditions required to obtain a targeted coating composition and can be used in the future as a starting point for the establishment of a full composition-microstructure-property chain. We demonstrate that this AI-assisted composition-prediction and deposition-optimization strategy performs accurately within the range of alloy systems investigated, paving the way toward autonomous multi-objective optimization of protective coatings for next-generation nuclear and other low-carbon energy technologies.

# I. Coatings for Carbon-free energy in PEPR-DIADEM

A main goal of contemporary research is to bring major innovation in the 3 main transitions the world has to face: energy, digital and healthcare transition to move towards decarbonized energy, digital transformation or new devices for healthcare technologies. These innovations are strongly related to Materials Science and Engineering and the need for new materials always more efficient and sustainable is pressing. This challenge does not only mean finding materials with desired properties, it requires to find the most relevant way of synthesis to develop the properties of a given material at its full potential. Theses ways can be using solid state synthesis or melting to obtain bulk materials, additive manufacturing to get materials with complex shapes, nanoparticles for wide surface interaction or thin films to control the interaction between two media.

The recent development of AI solutions has brought new opportunities to face these challenges and accelerate the discovery of new materials and the exploration of their properties. It impacts the whole process of materials science from synthesis (adaptation in real time of the synthesis conditions) to the structural, physical and chemical characterization of always more complex systems in more and more powerful setups, for example in large scale facilities like synchrotrons [1], and the evaluation of their performance for the desired application (corrosion, energy production, …). This trend started in 2011 with the Materials Genome Initiative [2], a research program led by the US government and the number of publications increased exponentially since. Other mile stones of this trend include the development of Machine Learning architecture applied to materials in 2018, the development of self-driving labs (A-labs) in 2020 and the Nobel prize in chemistry for computational conception of proteins (see Figure 1).

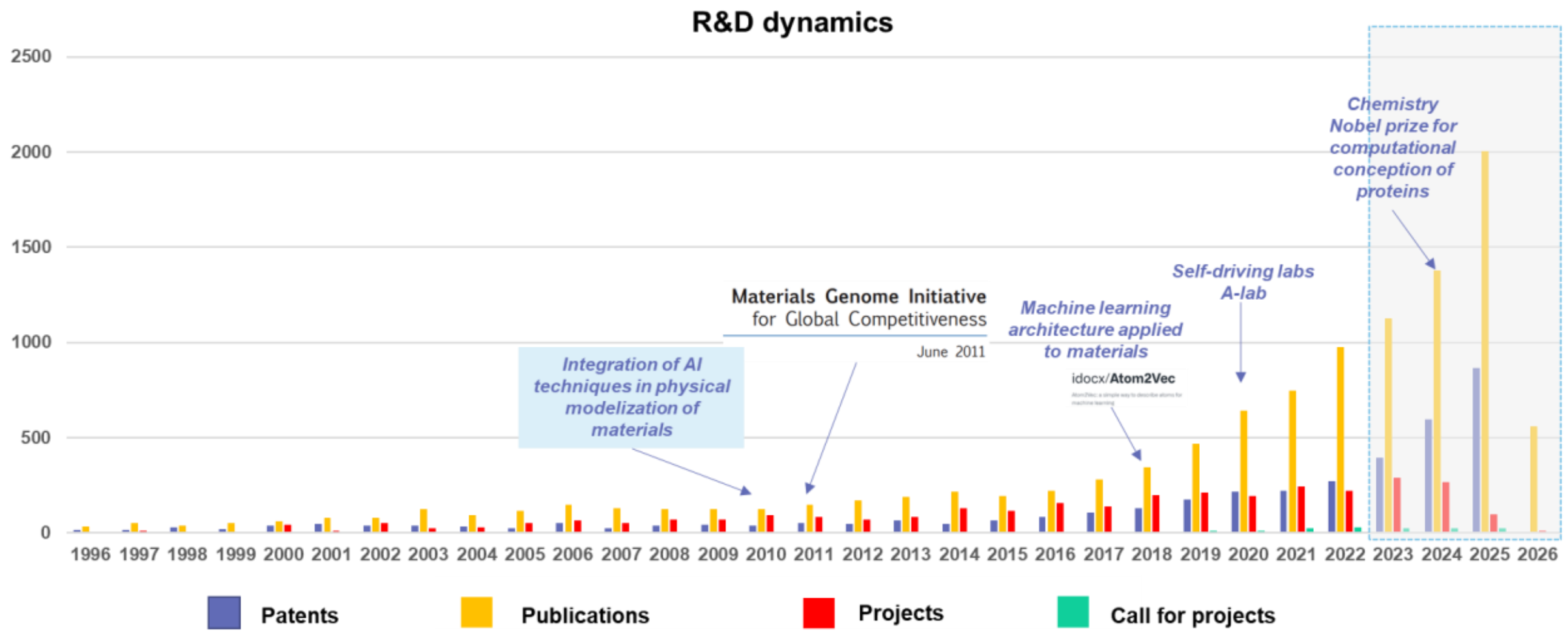

*Figure 1: Worldwide R&D dynamics in Data and Materials science convergence over a 30 years period starting in 1996 with the main milestones*

When a country-by-country comparison is made (see Figure 2), one can observe that the number of publications in the field increases sharply since 2016 (+28% annual growth) mostly led by China (+44% annual growth compared to +21% annual growth for Europe and USA).

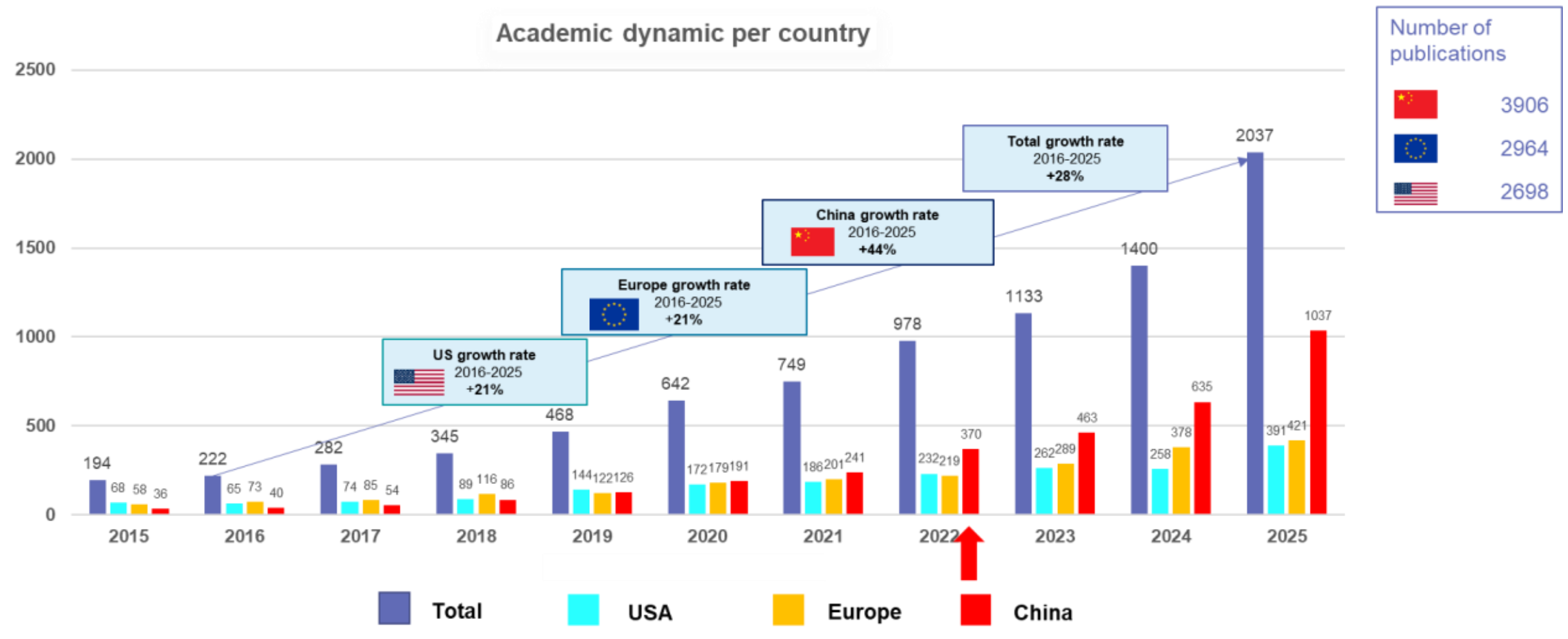

*Figure 2: Academics dynamics per geographical region over a 10 years period starting in 2015 showing an increase of China dynamics in number of publications starting in 2018 (red arrow).*

At the European level (see Figure 3), various national initiatives have been implemented like Material Digital [3] in Germany, AI for science strategy and Digital material foundry in the UK, the MI2I in Japan or the Acceleration Consortium in Canada leading to numerous and valuable international collaborations. To tackle this new paradigm in the way of doing research, the French national initiative **DIADEM** (**DI**scovery **A**cceleration for the **D**eployment of **E**merging **M**aterials) led by CEA and CNRS was launched in 2022, thanks to the dynamics it is creating and the organisation in this field of research at the national level led to a sharp increase of the number of publications in the 2022-2024 period (+150%).

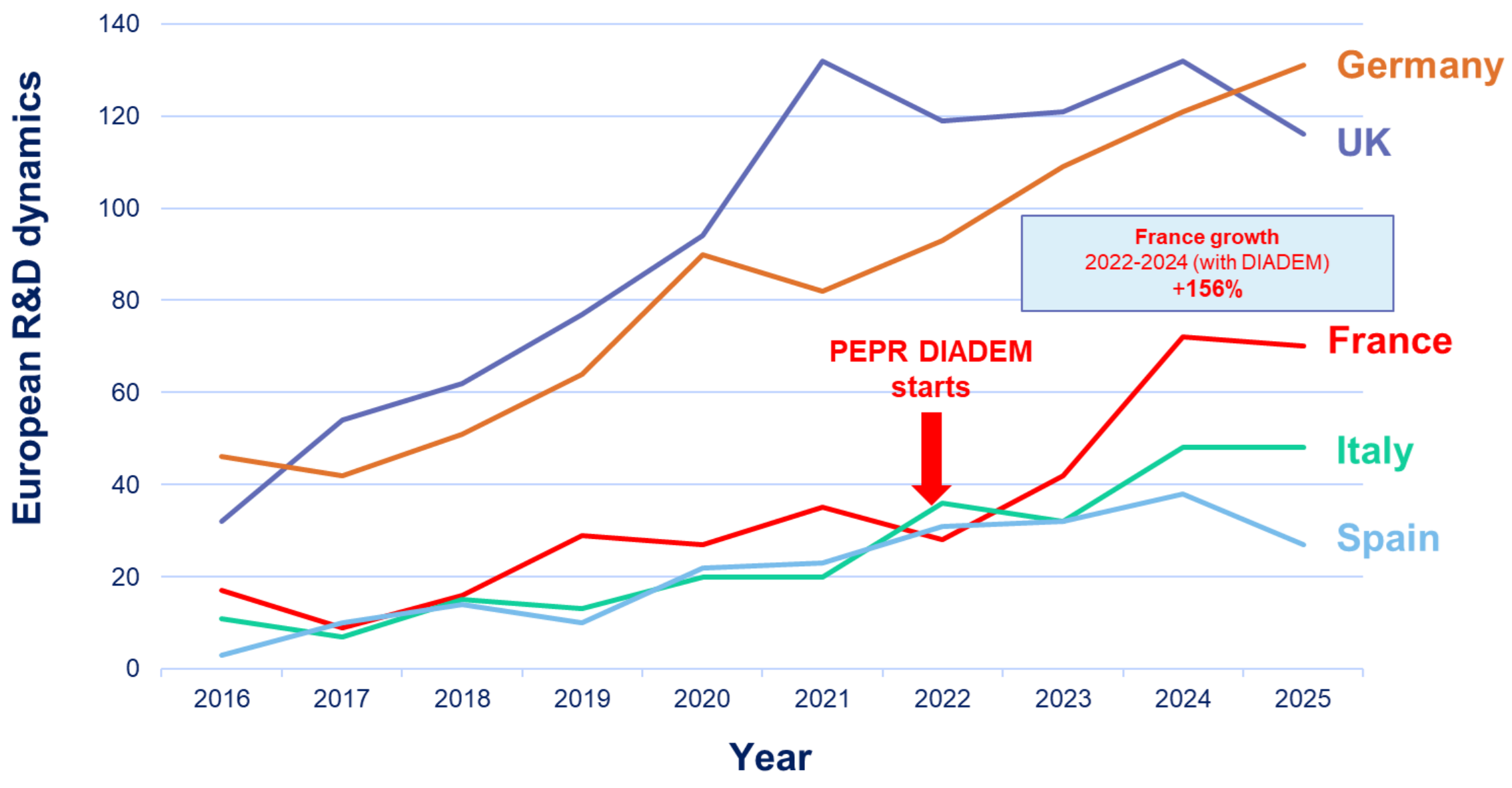

*Figure 3: European R&D dynamics in data and materials science convergence over a 10 years period starting in 2016. France dynamics in this field increases since 2022 thanks to PEPR DIADEM dynamics.*

The DIADEM initiative has been implemented aiming to organise at the national level the research on the convergence of Materials and Data science [4]. It targets a wide variety of applications, from nuclear energy to nanoelectronics and healthcare.

Thin films technologies are at the heart of developments in innovative materials for carbon-free energy, both for nuclear power of today and tomorrow (SMR) and also for the hydrogen production sector for example. To date, several flagship projects related to energy transition towards a decarbonized future are dedicated to the development of innovative coatings materials, especially for harsh environments:

- A-DREAM for corrosion resistant materials in Small Modular Reactors (especially resistant in molten salts media)
- ASTERIX for protective coatings for nuclear enhanced accident tolerant fuel
- CROCUS for new materials integrated in nuclear fuel reprocessing (in particular resistant in concentrated nitric acid media)
- COLLECTIF for the development of coatings to enhance the durability of high temperature electrolysers stacks
- DETOX explores electrochromic thin films
- PARADOX will focus on High Entropy Oxides thin films by combinatorial synthesis both for nuclear and hydrogens sectors
- MARINA studies materials for antibiofouling and corrosion resistance in marine environments

The main objective of these research projects is to use the network of acceleration platforms that has been established in France, and particularly the high-throughput PVD synthesis platforms such as the DIADEM 2D platform.

The majority of these projects rely on the use of high entropy materials (alloys, oxides). The former have attracted much attention recently [5], [6] due to their strong resistance to corrosion and mechanical properties [7], [8]. The main aspect of these materials made of five elements or more is the exploration of the central zone of phase diagrams as a single phased zone which exists in contradiction with Gibbs phase rule. This observation comes from the increase of the configuration entropy $\Delta S_{conf}$ with a maximum of the Boltzmann law reached when the concentration of the constitutive elements are equivalent [9]. This effect of entropy has been generalized recently to the case of high entropy ceramics [10] which crystallize in many crystallographic systems : rock salt [11], fluorite [12], spinel [13], pyrochlore [14], magneto-plumbite [15], perovskite [16], garnet [17] and rutile [18] with a wide variety of properties which make them interesting for different applications like thermoelectricity, magnetism, catalysis or batteries [19].

High entropy materials are subject to cocktail effect [20] stating that the material structure and properties are determined by the combination of the properties of the elements and the interactions between them [21] leading to the fact that a particular property of high entropy materials cannot be linked to a single element. This effect is expected to be at the origin of their exotic properties but is complex to understand due to the high complexity of these alloys. Consequently, the use of artificial intelligence which increased drastically in the recent years may contribute to a better understanding and determination of their properties and their ways of synthesis explaining why this family of materials is well represented among PEPR DIADEM projects.

## II. Experimental methods

During this study, new coatings have been produced using DIADEM-2D (see Figure 4), a combinatorial PVD platform located at CEA-INSTN-Saclay. This acceleration platform, dedicated to research and education, is fitted with 4 independent confocal cathodes – 2 HiPIMS [22] and 2 pulsed-DC – giving the ability to deposit either a target composition at the centre of the substrate holder or composition gradients using wider samples. In the latter situation, this leads to the introduction of composition high speed screening on this platform to accelerate material discovery.

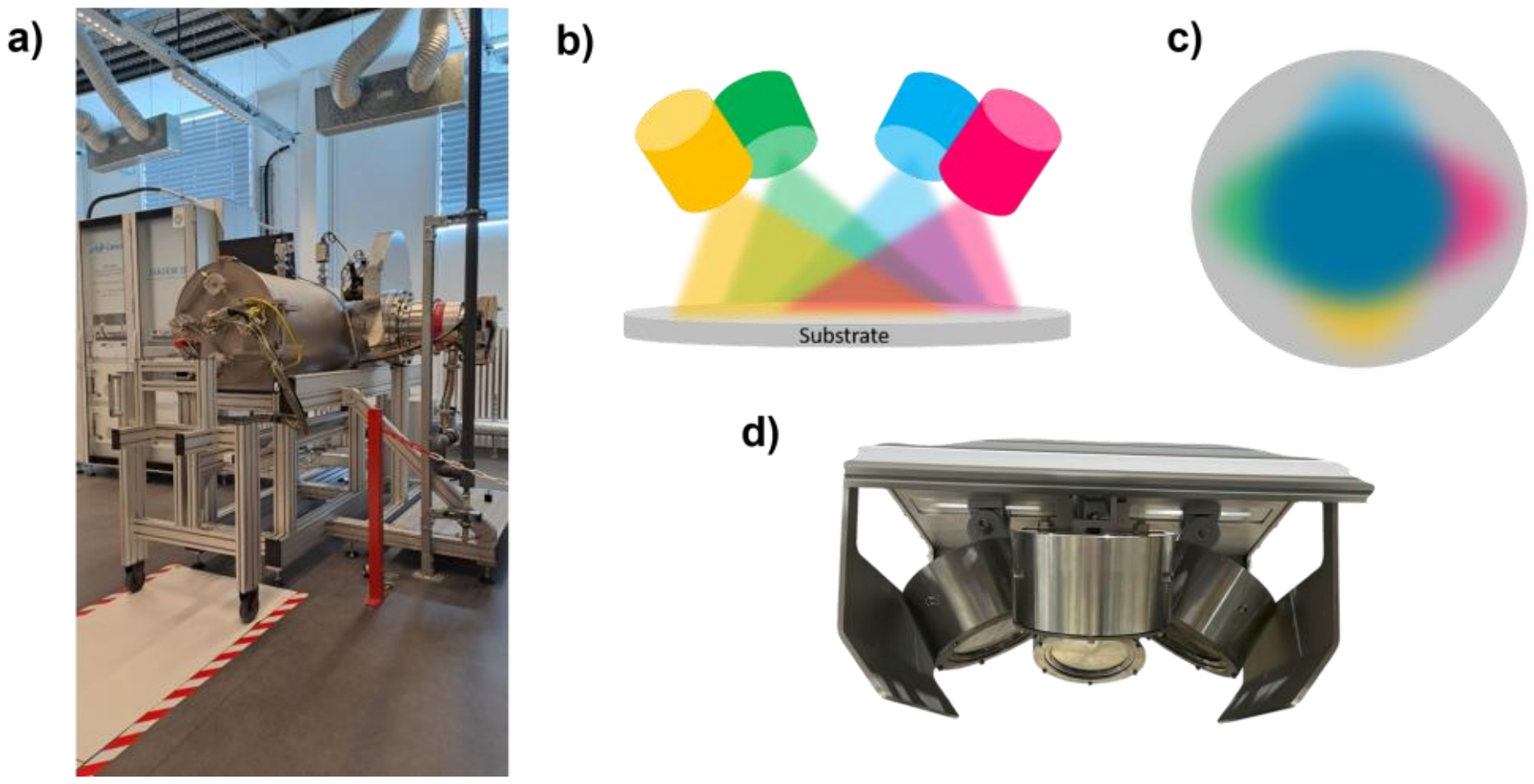

*Figure 4: a) DIADEM-2D platform at INSTN-Saclay, b) Confocal combinatorial deposition as fitted in DIADEM-2D, c) Deposition pattern, d) Multi-cathode system in DIADEM-2D*

This platform has been thought with the aim to become an autonomous thin film deposition platform. To this end, it is equipped with several in situ monitoring: Optical Emission Spectroscopy and Ion/neutral ratio to get real time feedback of the chemical composition and microstructure. To achieve this goal of autonomous deposition, this platform is using AI models to determine the targeted chemical composition prior to the deposition (detailed in III) using ex-situ and in-situ chemical characterizations as a database for AI models (Figure 5 a). However, the final goal is to reach the autonomous this film deposition using in-situ real time monitoring: the AI models would be able to improve in real time the growth condition to reach particular thin films compositions or properties (Figure 5 b).

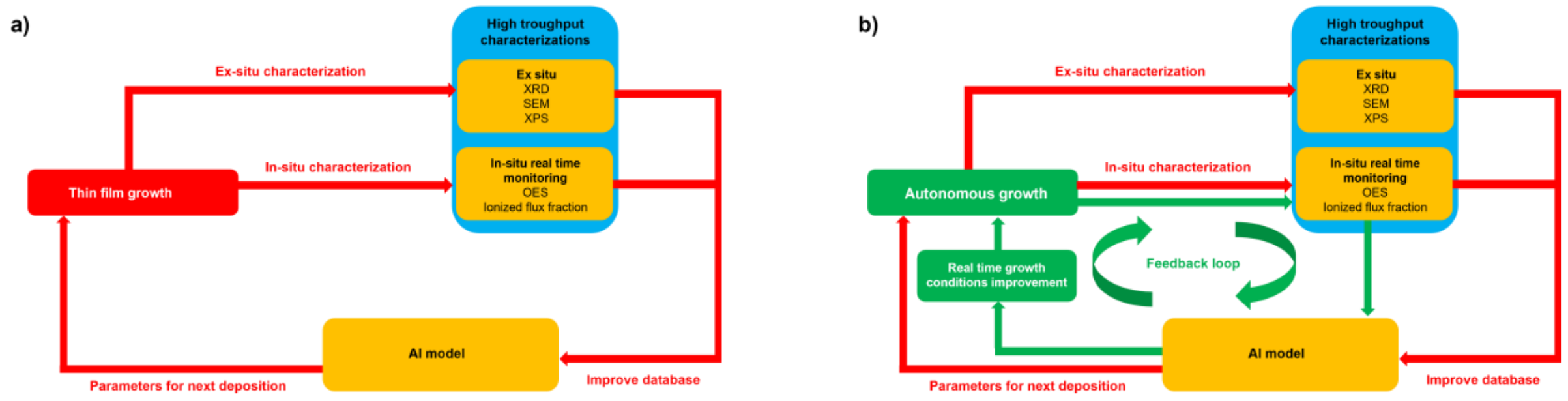

*Figure 5: Operation description of DIADEM-2D a) to date, b) in a near future including an AI-controlled feedback loop for autonomous thin film deposition*

In this first study, thin films are grown on p-doped 700 µm thick Si(100) substrates. Magnetron targets are made of a single element (Cr, Al, Mo, Ni, V, W, Nb) with 99.9% purity, the substrate is at room temperature during deposition. All samples have been cleaned through an etching process before deposition with -900V substrate bias: Diode etching at $10^{-2}$mbar Ar working gas pressure followed by Cr ion etching using HiPIMS cathode (impulsion time 30µs, period 920µs, 250W) for a total duration of 30 minutes. The deposition has been done using -100V bias with Ar working gas pressure of $5.10^{-3}$mbar (Ar gas debit of 40 SCCM), the constant parameters applied on HiPIMS cathodes are: 30µs impulsion time, period 920µs, the parameters applied on pulsed-DC cathodes: 4µs impulsion time, 50kHz frequency.

Thin film composition was measured using EDS spectra acquired under 15kV acceleration voltage in a JEOL IT500HR SEM-FEG with Bruker EDS spectrometer.

# III. Chemical system determination

The chemical system determination for each application was realized using data analysis from the literature.

## 3.1 ADREAM – corrosion in molten salt media

ADREAM project is exploring new materials resistant in molten salt media, mostly Small Modular Reactors (SMR). These materials can be either bulk or coatings, each having their specificities and requirements. Thin films give the opportunity to study new compositions easily: a new composition can be produced very quickly by combinatorial synthesis, and gives the opportunity to separate the chemical properties from the structural ones. Indeed, only the surface is interacting with the molten salt, the most efficient way to process is to find the material with the best chemical properties for the interface and the material with the best mechanical properties for the bulk.

The selection of the coating materials was based on their expected corrosion performance in molten salt environments. A previous analysis [23] using pairwise comparison and Spring Rank was used to rank alloys from literature and experimental corrosion data by assigning a corrosion performance score to each alloy (see Figure 6). This ranking was then used as input for a Gaussian Process model to establish the relationship between alloy chemical composition and corrosion performance. Based on this approach, the **Ni-Cr-Mo-Al** chemical system was identified as a promising candidate for protective coatings.

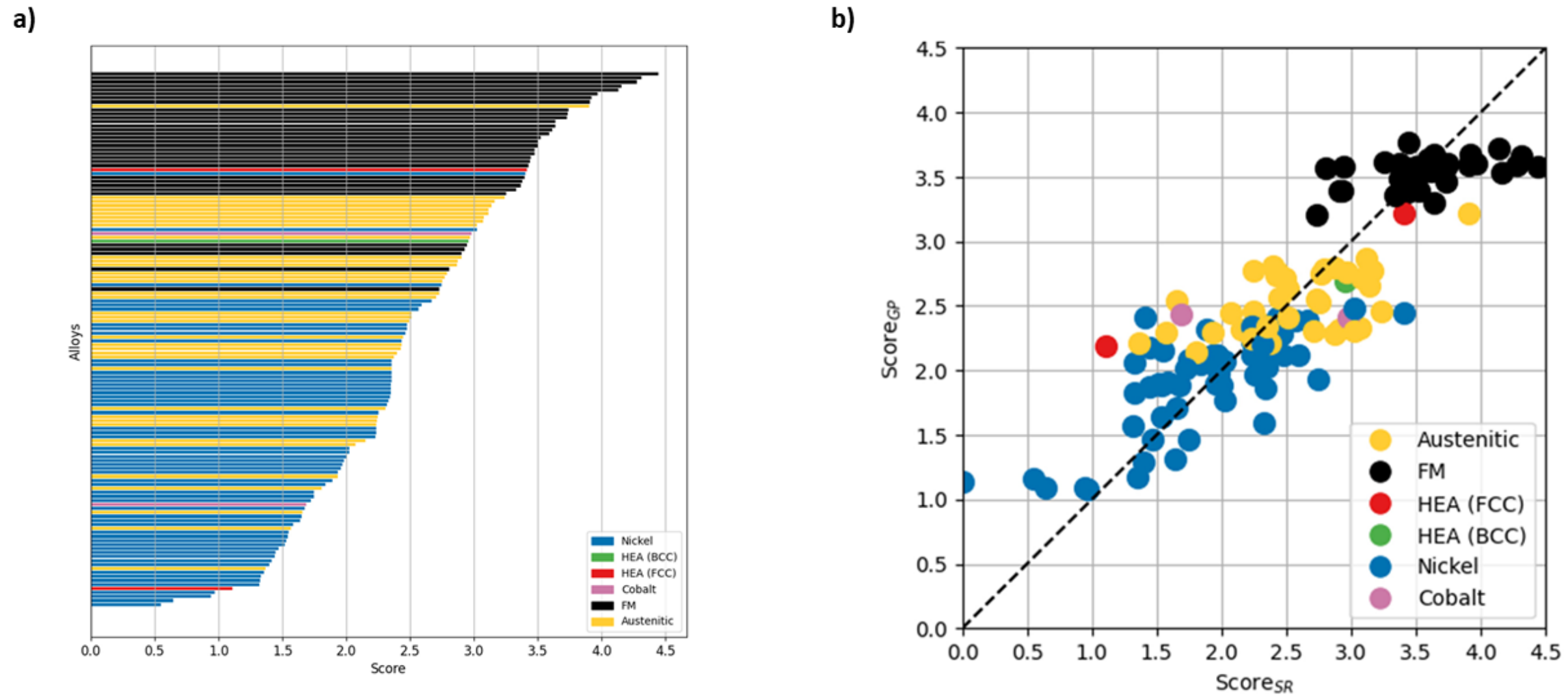

*Figure 6: a) Alloy ranking analysed in molten salt corrosion experiments, b) Comparison of scores computed by Spring Rank and Gaussian Process (adapted from [23])*

The relevance of this chemical system for molten-salt corrosion protection can be rationalized physically: Cr and Al both promote the formation of dense, protective oxide scales ($Cr_2O_3$- and $Al_2O_3$-based passive layers) that limit dissolution and ionic transport at the coating/salt interface, while Mo contributes solid-solution strengthening and further improves the intrinsic corrosion resistance of Ni-based alloys in halide environments [24]. These qualitative trends are consistent with the corrosion-performance ranking obtained from the pairwise-comparison/Spring Rank analysis and the Gaussian Process model (Figure 6), which were used here to select the target chemical system; a quantitative demonstration of the corrosion behaviour of the deposited coatings themselves is beyond the scope of the present study and is addressed in the outlook (see Conclusion).

## 3.2 ASTERIX – Coatings for nuclear claddings

ASTERIX project is exploring new coatings to increase the resistance to corrosion of nuclear claddings in accidental conditions. Previous studies developed enhanced accident tolerant fuel claddings with the deposition of a chromium layer onto nuclear cladding made of zirconium alloy by PVD-HiPIMS [25]. This process limits high temperature and dihydrogen production in the event of cooling shortage. In addition, it limits cladding creeping and bloating, preserving core geometry. However, the resilience of this process may be reduced by the strong solubility of chromium in β-Zr phase leading to a eutectic melting point at 1316°C (see Figure 7 a). In case of cooling shortage which could lead to overshoot this temperature, the existence of this eutectic point may induce cladding melting and warping which would be a serious hazard (see Figure 7 b).

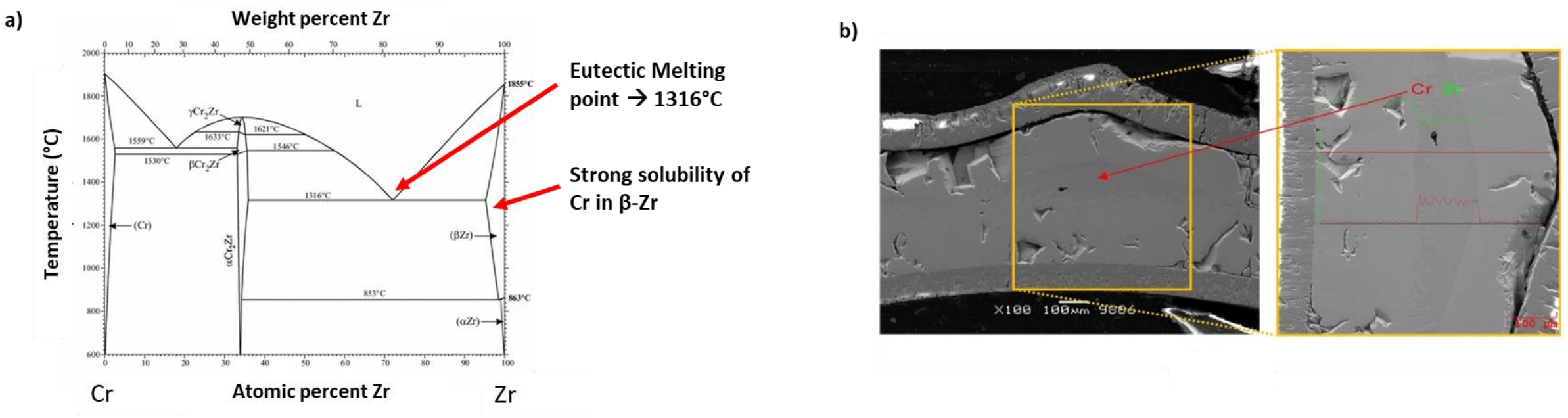

*Figure 7: a) Cr-Zr phase diagram, b) SEM image with EDS profile of a Cr-coated Zr alloy cladding displaying bloating and Cr miscibility in Zr alloy after heating at 1400°C for 2 minutes in steam environment, c)12-15µm Cr-coated M5$_{Framatome}$ nuclear cladding after HT oxidation for 100s at 1400°C followed by direct quenching. Adapted from [26] and [27]*

To limit this phenomenon, a solution would be to include a diffusion barrier between zirconium alloy and the chromium layer avoiding a direct contact between chromium and zirconium (see Figure 8 a). Knowing the extreme conditions this barrier would have to face, the best material would be refractory high entropy alloys coating thanks to their high melting point and strong corrosion resistance [28], [29]. To determine the best composition, a multicriteria optimization was realized (thermodynamics, diffusion, neutronics) in the global chemical system: Fe-Cr-Zr-Mo-V-Nb-W-Ti (see Figure 8 b). Realizing a Pareto front enabled to select the best diffusion barrier composition : V-Nb-Mo-W which has already been mentioned for this application [30]. Previous studies investigated the refractory alloy system Nb-Ti-V-Zr [31] which could be interesting for our application as well and using combinatorial synthesis but the disadvantage of not being exempt from Zr.

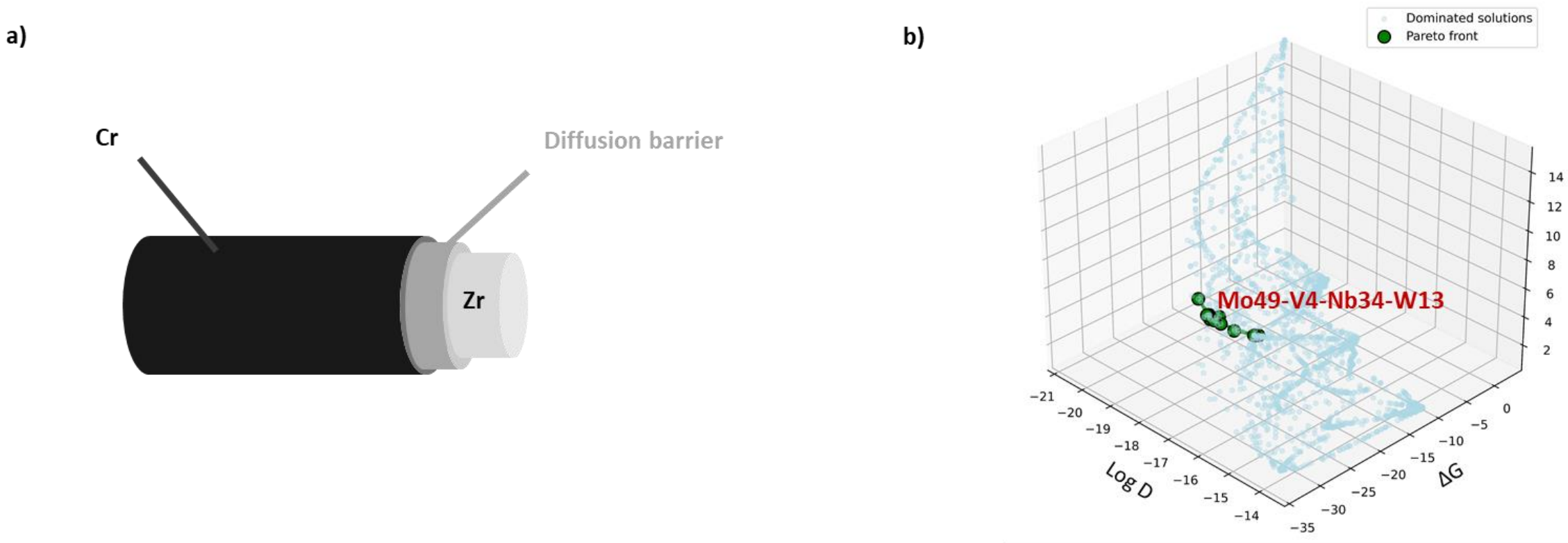

*Figure 8: a) Architecture of a Cr + diffusion barrier coated Zr alloy nuclear cladding as planned in the ASTERIX project, b) Pareto front from a multicriteria optimization using a thermodynamic database filled with data from the literature. The composition in red appears to be the best composition for the diffusion barrier*

Within the V-Nb-Mo-W system identified by the Pareto optimization, each element addresses a specific functional requirement of the diffusion barrier: V, Nb, Mo and W are refractory elements with high melting points and limited mutual solubility with Zr [32], which restricts interdiffusion and preserves barrier integrity at high temperature, while Nb and Mo further contribute favourable neutronic cross-sections and W maximizes thermal stability [33], [34]. This composition therefore represents a compromise identified simultaneously from thermodynamic, diffusion and neutronic criteria (Figure 8b), rather than an optimum for a single property. Experimental validation of the diffusion resistance and thermal stability of this composition, once deposited as a thin film, is planned as a follow-up to the present work (see Conclusion).

# IV. Artificial Intelligence for thin film optimization

## 4.1. Context

Combinatorial deposition is a powerful tool to obtain thin films with a particular composition and microstructure but the process to achieve these goals of thin film properties can be time consuming in a trial-and-error methodology because of the existence of a high number of tuneable parameters. For example, the deposition parameters that can be tuned include the power on each cathode, HiPIMS pulse duration, on/off time ratio, Ar pressure, bias voltage, … To tackle this issue, a powerful way is to use artificial intelligence to sort correlations between parameters in order to make a prediction of the thin film characteristics (chemical, physical, …) from the deposition parameters and, even more interesting, the deposition parameters to use in order to reach particular properties.

Artificial intelligence can be sorted in different categories, from algorithms where human action is mandatory to deep learning like neural networks in which the program acts by itself. Between these two extremes, machine learning models like XGBoost and random forest are relevant for the aim targeted in this publication as it works fine with reduced data set (≈100 datapoints) comparing to neural network for which dataset containing much more datapoints are required (≈1000 datapoints). In addition, they can be much more monitored than neural networks with hyperparameters determined by the user

## 4.2. An element-independent model

Machine learning models were developed to learn both the direct relationship between the sputtering powers applied to the cathodes and the resulting thin-film composition, and the inverse relationship allowing the determination of the required powers to obtain a target composition. To fully exploit the experimental data generated by the combinatorial deposition process, we further developed a generic, element-independent modelling framework. Instead of relying on the chemical identity of the deposited elements, the model uses descriptors based on the deposition configuration (e.g., cathode operating mode) together with normalized physical parameters of each element, such as the sputtering yield and the atomic mass. By encoding the deposition process through these physical descriptors, the same model can be applied to different alloy systems without requiring the acquisition and training of a dedicated dataset for each new material combination. This approach greatly reduces the experimental effort associated with model development while considerably improving the flexibility, transferability, and productivity of combinatorial sputtering studies. It also enables rapid adaptation to new material systems or to modifications of existing ones, such as the replacement or addition of an element, with only limited additional experimental data.

The choice of these descriptors is physically motivated. The sputtering yield of each target element governs, for a given cathode power, the atomic flux reaching the substrate, which directly controls the local supersaturation and the resulting film composition. The atomic mass is linked to the kinetic energy and is linked to the energy of the particles have when reaching the substrate and to the way the fluxes from the cathodes interact. The deposition-configuration descriptor (HiPIMS vs. pulsed-DC operating mode) encodes the ionized fraction of the sputtered flux, which affects ion bombardment energy and adatom mobility at the growing film surface and therefore the plasma-assisted alloy formation process. By combining these physically grounded descriptors rather than the raw elemental identity, the model is intended to remain applicable when an alloying element is replaced or added, provided the physical descriptors of the new element are known, without requiring a full dedicated dataset for each new combination.

It should be noted, however, that in the present study this framework has been validated only in an interpolative regime, using a single dataset of 82 experimental points spanning seven elements (Ni, Cr, Mo, Al, W, Nb, V). The reported accuracy therefore reflects the ability of the model to predict compositions and process parameters within this compositional space, rather than a demonstrated capability to extrapolate to entirely unseen elements. Extending the validation to elements absent from the training set is identified as a priority for future work (see Conclusion).

## 4.3. Calculation details

The calculation environment was developed in python environment using scikit-learn package [35] for Random Forest [36] and XGBoost [37]. The dataset was split randomly into a train dataset (80% of the original dataset) and a test dataset (20% of the original dataset). Then the hyperparameters of each model have been optimized using Optuna optimizer [38]. The best finetuned model is used to get a value of composition (powers) from a set of powers (composition). To increase the relevance of our predictions, we implemented cross validation: This process is repeated several times (typically 5) taking each time different randomly split train and test datasets giving each time a prediction with the same input. This gives an information about the reliance of the model (using the standard deviation of the chosen performance metric). The prediction that will be used in this study is the average of these five values (see Figure 9) [39].

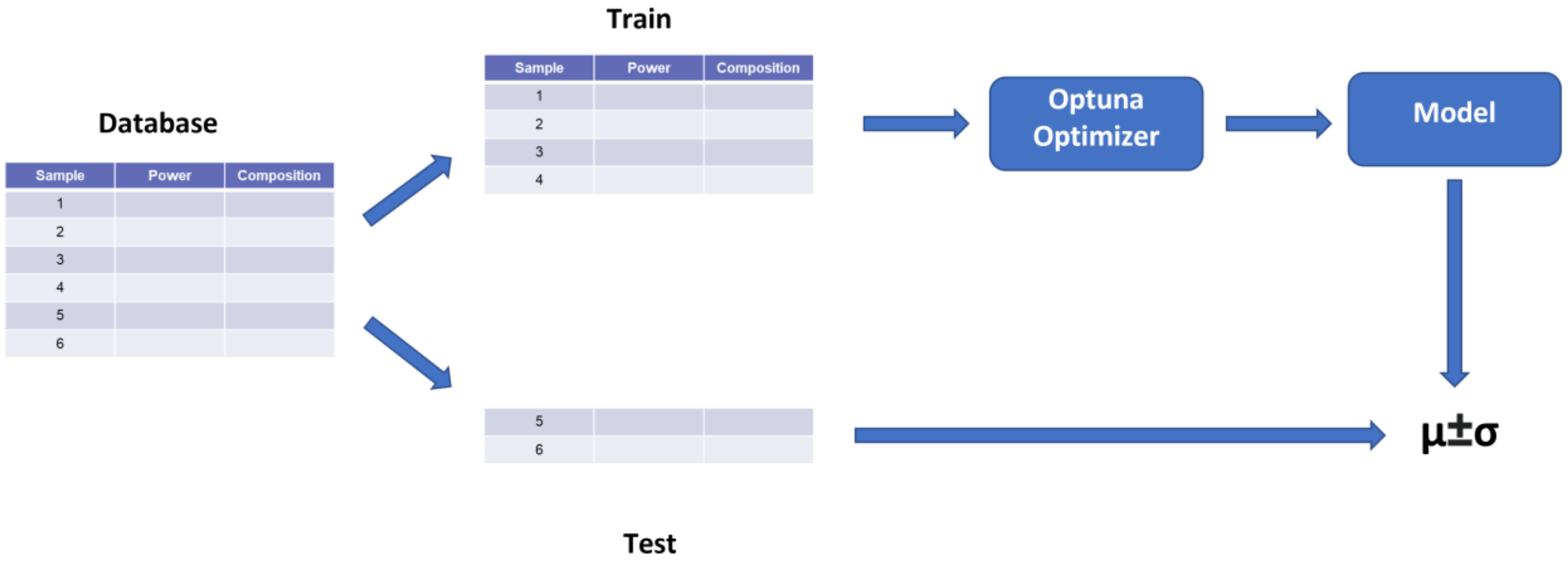

*Figure 9 : Sketch of the learning process used in this study using test and train batches, optuna optimizer and cross validation*

## 4.4. Predictions

The quality of the prediction of a parameter can be done comparing the predictions to corresponding experimental values in a test/train graph. We present in Figure 10 the train/test graph where was used a database of 7 elements (Ni/Cr/Mo/Al/W/Nb/V) containing 82 datapoints to predict the composition giving in input the power on the four cathodes (forward direction).

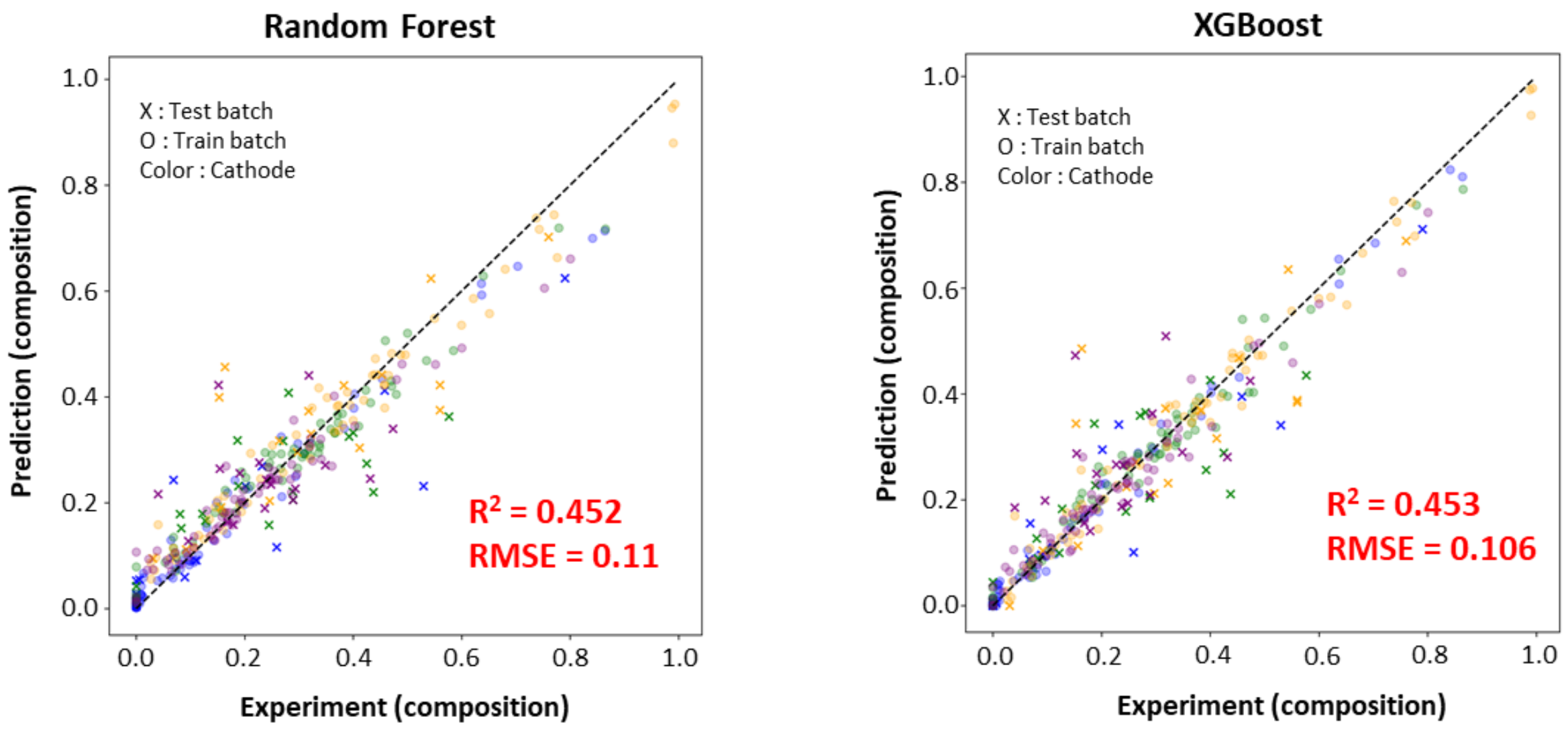

*Figure 10: Test/train batch using the element-independent model with Random Forest and XGBoost in the forward direction*

These graphs indicate that we have a fairly good prediction of the composition of the film as points appear to be close to the y=x line with RMSE≈0.1 which is making this way of working promising for thin film properties and characteristics prediction. We could certainly improve these curves adding more datapoints.

While the "forward" direction (power → composition) has a bijective character, the "backward" direction (composition → power) does not have this character as an infinite combination of powers could lead to the same thin film composition making the process in this direction more complicated as the learning may diverge more easily.

To overcome this difficulty, it is necessary to indicate windows in which the power can be predicted to limit divergence. As several models cannot predict values above or below output values in the dataset, we use a simple window to bound the values. Another strategy could be to limit the divergence at each step or to use a loop in the forward direction with an objective of composition, the loop would stop when the desired composition is reached.

We present in Figure 11 the test/train graphs in the "backward" direction using random forest and XGBoost.

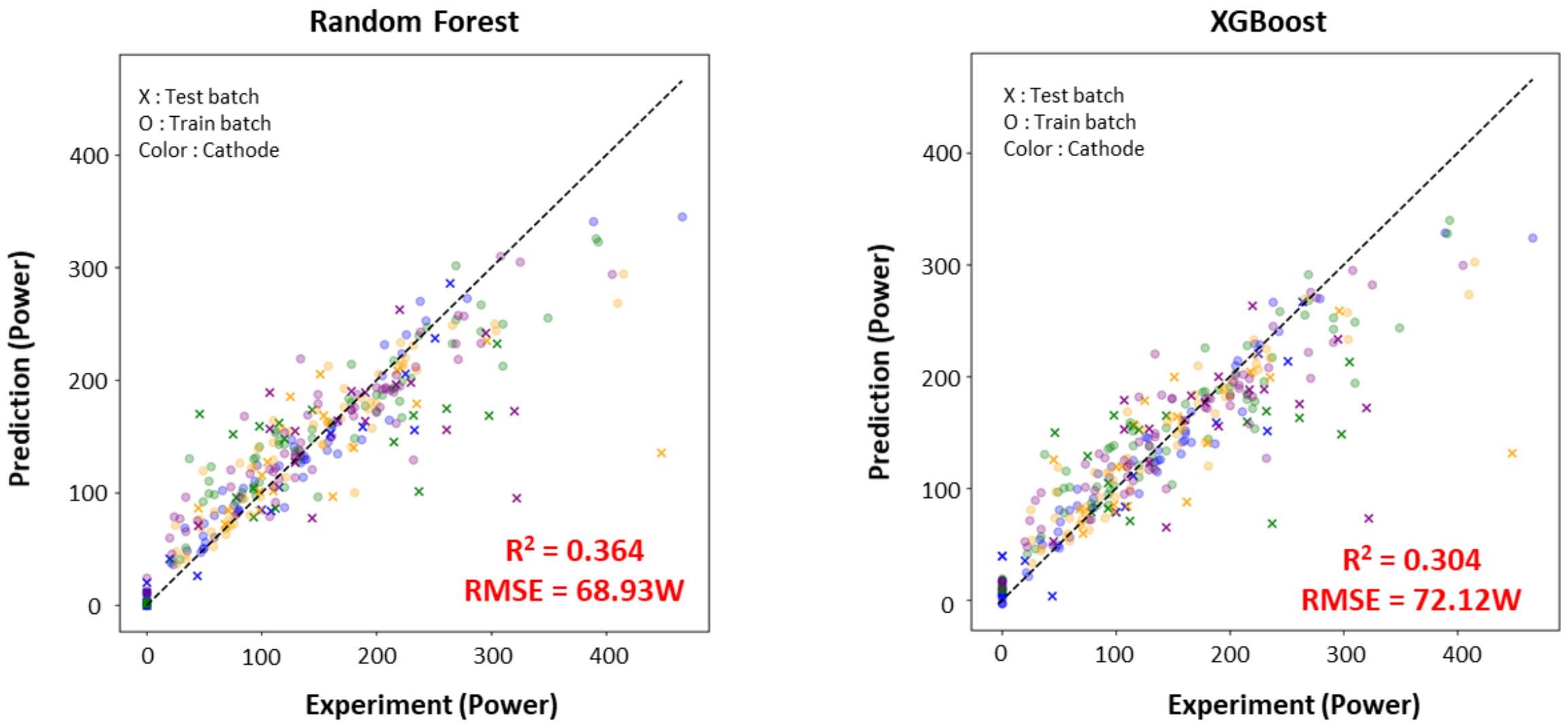

*Figure 11: Test/train batch using the element-independent model with Random Forest and XGBoost in the backward direction*

We obtain data points close to the y=x line with RMSE≈70W. These results show that we can use this methodology as a guide for process optimization with a limited dataset but the acquisition of more datapoints is necessary in order to improve the predictions of the model and going toward an effective prediction of process parameters. In addition, it shows the limitations that state-of-the-art AI models can suffer from. This methodology can be applied to dataset with elements which are not present in the dataset provided the physical deposition parameters are given in the database.

The systematically lower accuracy of the backward direction compared to the forward direction, observed consistently across all three modelling approaches (Random Forest, XGBoost and the fuzzy inference system, Section 4.6, is a direct consequence of the non-bijective nature of the composition-to-power mapping: several distinct combinations of cathode powers can yield the same thin-film composition, whereas each combination of powers yields, by construction, a single composition. This intrinsic asymmetry, rather than a deficiency of any specific algorithm, explains why the backward problem is more prone to divergence and why bounding the prediction window, as described above, is necessary to obtain physically meaningful results.

## 4.5. Correlation matrices

Another method to evaluate the learning quality of our models is to calculate correlation matrices and to compare them with experimental ones extracted directly from the database.

We show in Figure 12 the experimental correlation matrices in the “backward” direction where can be observed the strong correlation between the composition in an element and the power on the corresponding cathode as one would expect. However, the composition in one element appears not to be fully independent from the power on the other cathodes, evidence of the interaction between the plasma coming from the 4 cathodes.

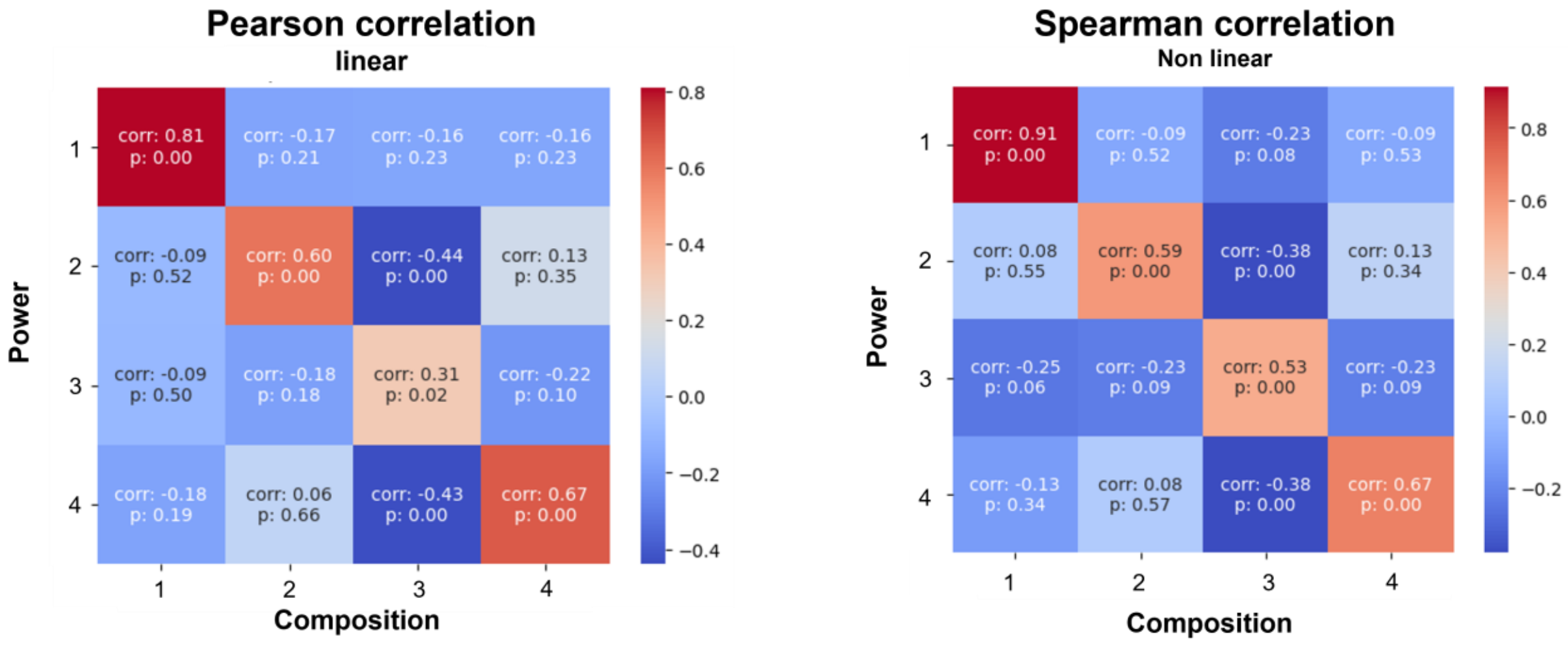

*Figure 12: Pearson and Spearman experimental correlation matrices calculated using data from the experimental database*

We used 3 machine learning techniques (Random Forest, XGBoost and Bayesian ridge [40], [41]) to create a model from the dataset and extract the correlations, we show the result in Figure 13. One can observe a fairly good agreement between experimental correlation matrices and predicted correlation matrices showing that these models can predict how parameters influence each other.

This inter-cathode coupling is expected physically, since the four confocal cathodes share the same deposition chamber and their respective plasmas partially overlap above the substrate; increasing the power on one cathode locally modifies the working-gas pressure and the ion/neutral flux ratio experienced by the other three, which explains why the composition associated with one cathode is not fully independent from the power applied to the others. The good agreement between the experimental correlation matrices (Figure 12) and those reconstructed from the model predictions (Figure 13) indicates that Random Forest, XGBoost and Bayesian ridge all capture this coupling to a similar extent, which supports their use for process optimization despite their differences in raw prediction accuracy (Section 4.4).

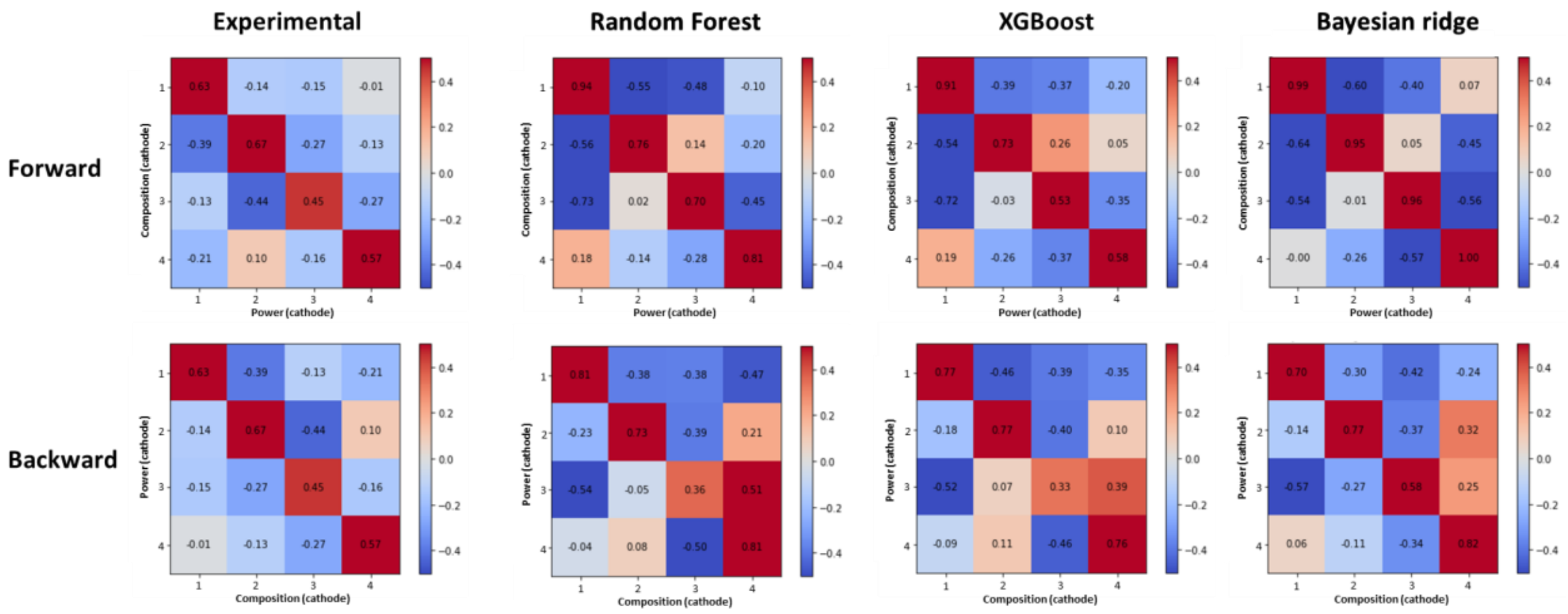

*Figure 13 : Forward and backward correlation matrices using experimental data and predictions using Random forest, XGBoost and Bayesian ridge*

Finally, we show in Figure 14 the training Root Mean Square Error (RMSE) which evaluates the quality of the training and the Coefficient Variation Root Mean Square Error (CV RMSE) evaluating the predictive quality of the model. In order to have a good learning process, these indicators must converge. It is the case in our study, then we can say that the results shown in this article come from models which learnt correctly from the database.

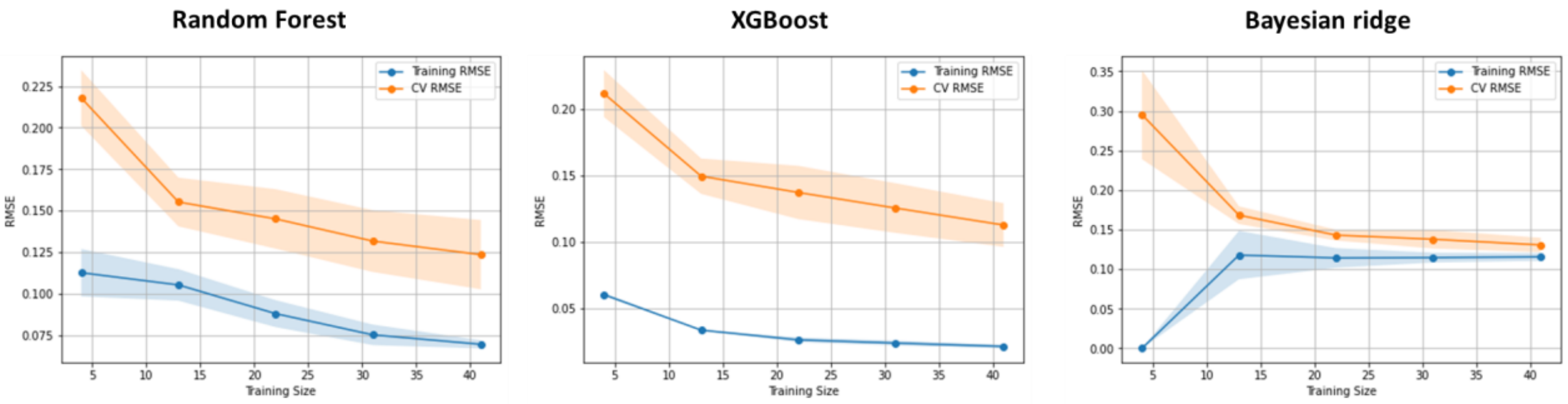

*Figure 14 : Learning curves displaying training RMSE and CV RMSE for the calculations of Random Forest, XGBoost and Bayesian Ridge models*

## 4.6. Leveraging fuzzy rule base systems

As showed in the previous section, state-of-the-art AI models struggle with the experimental data in different ways. Either the model needs a lot of data, which is more compatible with simulated data than experimental one, or they are considered as lower in terms of prediction (like tree-based ensemble methods, e.g. Random forests and XGBoost). Moreover, the most frequently used models in Materials Science literature are not transparent models. That implies researchers cannot get information from them. DIADEM also allows to investigate transparent models, like fuzzy systems, as proposed in [42], [43]. They offer good prediction performances while being more transparent. In DIADEM, they are still investigated to be improved, both in term of prediction and interpretability.

We tested fuzzy inference systems (FIS) with the algorithm described in [43] with the same protocol as presented before. In the forward direction, the approach got a $R^2 \approx 0.52$ and RMSE $\approx 0.106$. This slightly outperforms XGBoost (at least for $R^2$). In the backward direction, the same approach got a $R^2 \approx 0.24$ and RMSE $\approx 79.5$W. This is lower than the other models (regarding $R^2$, not RMSE) because the approach is more sensible to surjectivity. The problem of exploration is not as straightforward as we can imagine, and must be tackled by a smarter exploration approach. Figure 15 shows the results in both directions.

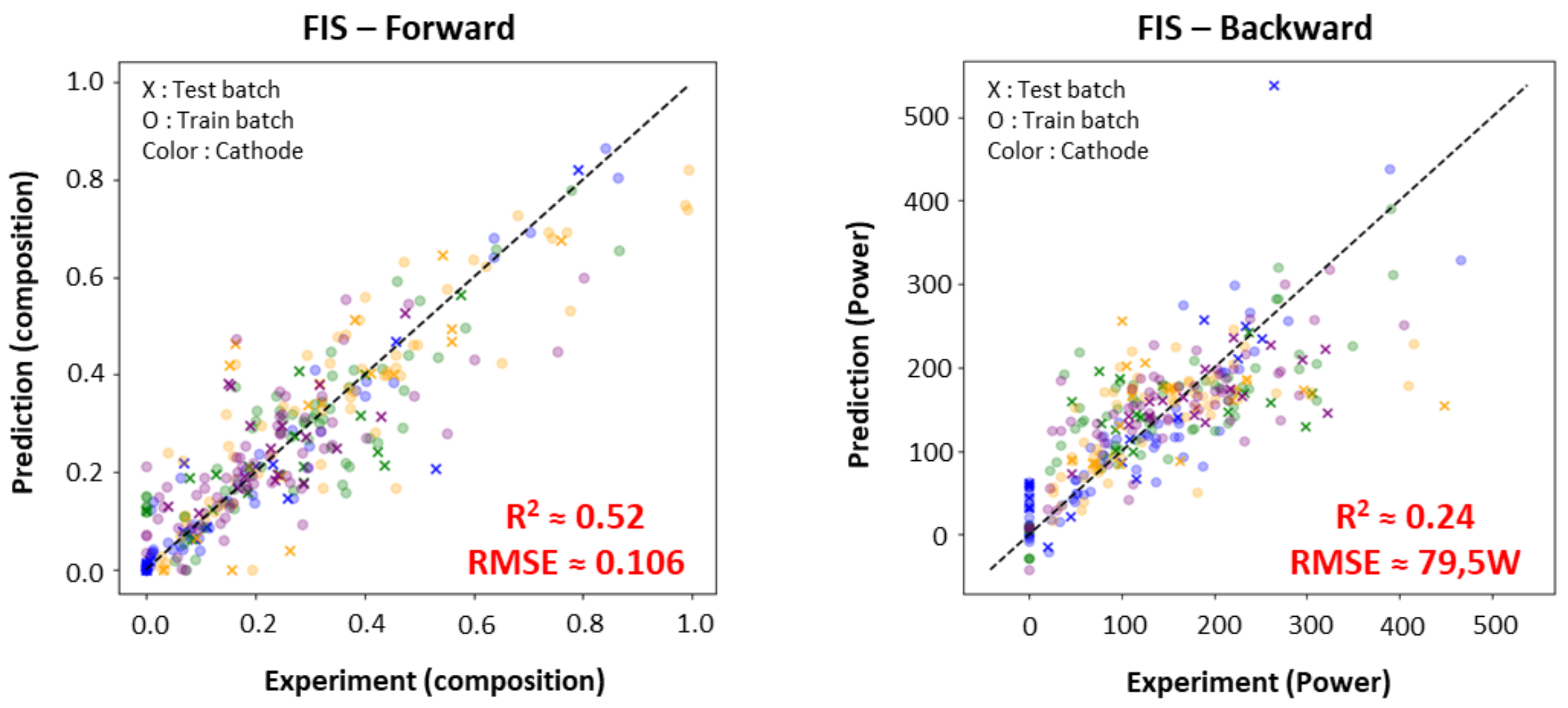

*Figure 15: Test/train batch using the element-independent model with a Fuzzy Inference System. On the left-hand side, the forward problem and on the right-hand side, the backward problem.*

The higher $R^2$ obtained with the fuzzy inference system in the forward direction, compared to XGBoost, together with its lower performance in the backward direction, mirrors the same asymmetry discussed in Section 4.4: fuzzy rule-based systems are particularly effective at capturing smooth, monotonic relationships such as the forward mapping from power to composition, but are more sensitive to the loss of surjectivity that characterizes the backward problem. This trade-off, combined with the transparency of the resulting rule base, motivates the continued development of fuzzy approaches within DIADEM as a complement to, rather than a replacement for, tree-based ensemble methods.

# Conclusion

This work demonstrates a first prototype of a descriptor-based, element-agnostic machine learning framework capable of predicting the PVD process parameters required to obtain a target coating composition, and of solving the corresponding inverse problem from process conditions to composition. Using Random Forest, XGBoost and a fuzzy inference system trained on 82 experimental data points spanning seven elements, the forward (power-to-composition) prediction reaches $R^2 \approx$ 0.45–0.52 and RMSE ≈ 0.11, while the backward (composition-to-power) prediction reaches $R^2 \approx$ 0.24–0.36 and RMSE ≈ 69–79 W, the lower backward accuracy reflecting the intrinsically non-bijective nature of that mapping rather than a limitation specific to any one algorithm. By relying on physically meaningful descriptors — sputtering yield and deposition configuration — rather than on the chemical identity of the deposited elements, the proposed approach represents a step toward transferable models for combinatorial thin-film deposition. We stress, however, that this validation remains interpolative within the seven-element dataset used here; extrapolation to entirely unseen elements has not yet been demonstrated and is identified as a priority for future work, together with the acquisition of additional experimental data to improve model accuracy.

The chemical systems selected in Section III (Ni-Cr-Mo-Al for molten-salt corrosion protection, V-Nb-Mo-W for high-temperature diffusion barriers) illustrate how literature-derived and experimental performance data can guide composition choices ahead of deposition; establishing a quantitative link between the deposition parameters predicted here, the resulting coating microstructure, and its functional performance (oxidation resistance, corrosion resistance, thermal stability, mechanical behaviour) remains the central objective of the next phase of this work.

Beyond the modelling framework itself, this methodology directly supports several industrial application areas within the scope of the PEPR DIADEM projects introduced in Section I: corrosion-resistant coatings for Small Modular Reactor components and molten-salt systems (A-DREAM), diffusion barriers for enhanced accident-tolerant nuclear fuel claddings (ASTERIX), protective coatings for nuclear fuel reprocessing equipment (CROCUS) and high-temperature electrolyser stacks (COLLECTIF), as well as, more broadly, functional thin films for marine antibiofouling and corrosion protection (MARINA) and other aerospace, catalysis and energy-related coating applications where rapid, data-driven composition and process optimization can shorten development cycles.

Future work will focus on three complementary directions. First, incorporating additional descriptors that govern coating performance — grain size, crystallographic texture, phase constitution, residual stresses, surface morphology and deposition rate — to progressively connect deposition conditions to coating microstructure and function. Second, constructing a comprehensive materials library for corrosion-resistant coatings, characterized through high-throughput localized electrochemical techniques coupled with operando Raman spectroscopy, to enrich the database and iteratively refine the model. Third, experimentally validating the diffusion-barrier compositions identified by multi-objective optimization through diffusion-couple testing, using PVD process parameters predicted by the present framework, to provide a complete demonstration of an AI-assisted inverse design workflow from targeted functional properties to experimental synthesis.

Ultimately, the combination of combinatorial PVD, high-throughput characterization and physics-informed machine learning opens the way to autonomous materials discovery platforms capable of identifying optimized coating compositions with reduced experimental effort, within the scope — and limitations — discussed above.

# Acknowledgments

The authors want to thank the *Agence Nationale de la Recherche* (National Agency for Research) of the French state for financial support under grants awarded by the PEPR DIADEM: ADREAM (ANR-22-PEXD-0003) and ASTERIX (ANR-23-PEXD-0004) and the CEA for co-funding via the cross-cutting program on Materials and Processes for financial support under grants Paradox, Detox and Marina and through the IMPACT chair of INSTN. The authors would like to thank Clotaire Chevalier and Bertrand Reynier for fruitful discussions and help for SEM-EDS measurements.

# Author contributions

P. Foulquier: thin film deposition and characterization, data curation, calculations, conceptualization, methodology, writing – original draft, writing – review and editing. R. Haddad: thin film characterization, writing – review and editing. A. Mahmoud: Calculations. E. Monsifrot: DIADEM-2D design and assembly. F. Balbaud-Célérier: thin film characterization, conceptualization. J-Ph. Poli: calculations, conceptualization, writing – review and editing. F. Schuster: thin film deposition and characterization, data curation, calculations, conceptualization, methodology, writing – original draft, writing – review and editing.

# Data availability

Data are available from the corresponding author upon reasonable request.

[1] B. De Goes Foschiani, S. Bongiorno, O. Robach, O. Ulrich, J. Micha, et J. Eymery, « Multimodal Imaging of Strain and Light Emission of Core–Shell InGaN/GaN Wires under a Sub-Micrometer Polychromatic X-Ray Probe », *Adv. Mater. Interfaces*, vol. 13, nº 4, p. e01025, févr. 2026, doi: 10.1002/admi.202501025.
[2] J. J. De Pablo, B. Jones, C. L. Kovacs, V. Ozolins, et A. P. Ramirez, « The Materials Genome Initiative, the interplay of experiment, theory and computation », *Curr. Opin. Solid State Mater. Sci.*, vol. 18, nº 2, p. 99-117, avr. 2014, doi: 10.1016/j.cossms.2014.02.003.
[3] B. Bayerlein *et al.*, « Concepts for a Semantically Accessible Materials Data Space: Overview over Specific Implementations in Materials Science », *Adv. Eng. Mater.*, vol. 27, nº 8, p. 2401092, avr. 2025, doi: 10.1002/adem.202401092.
[4] F. Lomello, L. Bard, M. Maglione, et F. Schuster, « PEPR DIADEM: Priority equipment and research program on the development of innovative materials using artificial intelligence », *Comput. Struct. Biotechnol. J.*, vol. 25, p. 186-193, déc. 2024, doi: 10.1016/j.csbj.2024.09.019.
[5] B. Cantor, I. T. H. Chang, P. Knight, et A. J. B. Vincent, « Microstructural development in equiatomic multicomponent alloys », *Mater. Sci. Eng. A*, vol. 375-377, p. 213-218, juill. 2004, doi: 10.1016/j.msea.2003.10.257.
[6] J.-W. Yeh, « Alloy Design Strategies and Future Trends in High-Entropy Alloys », *JOM*, vol. 65, nº 12, p. 1759-1771, déc. 2013, doi: 10.1007/s11837-013-0761-6.
[7] Y. Y. Chen, T. Duval, U. D. Hung, J. W. Yeh, et H. C. Shih, « Microstructure and electrochemical properties of high entropy alloys—a comparison with type-304 stainless steel », *Corros. Sci.*, vol. 47, nº 9, p. 2257-2279, sept. 2005, doi: 10.1016/j.corsci.2004.11.008.
[8] Y. Y. Chen, U. T. Hong, H. C. Shih, J. W. Yeh, et T. Duval, « Electrochemical kinetics of the high entropy alloys in aqueous environments—a comparison with type 304 stainless steel », *Corros. Sci.*, vol. 47, nº 11, p. 2679-2699, nov. 2005, doi: 10.1016/j.corsci.2004.09.026.
[9] J. -W. Yeh *et al.*, « Nanostructured High-Entropy Alloys with Multiple Principal Elements: Novel Alloy Design Concepts and Outcomes », *Adv. Eng. Mater.*, vol. 6, nº 5, p. 299-303, mai 2004, doi: 10.1002/adem.200300567.
[10] C. M. Rost *et al.*, « Entropy-stabilized oxides », *Nat. Commun.*, vol. 6, nº 1, p. 8485, sept. 2015, doi: 10.1038/ncomms9485.
[11] S. S. I. Almishal *et al.*, « Untangling individual cation roles in rock salt high-entropy oxides », *Acta Mater.*, vol. 279, p. 120289, oct. 2024, doi: 10.1016/j.actamat.2024.120289.
[12] L. Su *et al.*, « Visualizing the Formation of High-Entropy Fluorite Oxides from an Amorphous Precursor at Atomic Resolution », *ACS Nano*, vol. 16, nº 12, p. 21397-21406, déc. 2022, doi: 10.1021/acsnano.2c09760.
[13] G. H. J. Johnstone *et al.*, « Entropy Engineering and Tunable Magnetic Order in the Spinel High-Entropy Oxide », *J. Am. Chem. Soc.*, vol. 144, nº 45, p. 20590-20600, nov. 2022, doi: 10.1021/jacs.2c06768.
[14] Z. Teng *et al.*, « Synthesis and structures of high-entropy pyrochlore oxides », *J. Eur. Ceram. Soc.*, vol. 40, nº 4, p. 1639-1643, avr. 2020, doi: 10.1016/j.jeurceramsoc.2019.12.008.
[15] D. A. Vinnik *et al.*, « High-entropy oxide phases with magnetoplumbite structure », *Ceram. Int.*, vol. 45, nº 10, p. 12942-12948, juill. 2019, doi: 10.1016/j.ceramint.2019.03.221.
[16] J. Ma, T. Liu, W. Ye, Q. He, et K. Chen, « High-entropy perovskite oxides for energy materials: A review », *J. Energy Storage*, vol. 90, p. 111890, juin 2024, doi: 10.1016/j.est.2024.111890.
[17] Y. Feng *et al.*, « Discovery of high entropy garnet solid-state electrolytes via ultrafast synthesis », *Energy Storage Mater.*, vol. 63, p. 103053, nov. 2023, doi: 10.1016/j.ensm.2023.103053.
[18] X. Miao, Z. Peng, L. Shi, et S. Zhou, « Insulating High-Entropy Ruthenium Oxide as a Highly Efficient Oxygen-Evolving Electrocatalyst in Acid », *ACS Catal.*, vol. 13, nº 6, p. 3983-3989, mars 2023, doi: 10.1021/acscatal.2c06276.
[19] S. Sen, M. Palabathuni, K. M. Ryan, et S. Singh, « High Entropy Oxides: Mapping the Landscape from Fundamentals to Future Vistas: Focus Review », *ACS Energy Lett.*, vol. 9, nº 8, p. 3694-3718, août 2024, doi: 10.1021/acsenergylett.4c01129.

[20] J.-W. Yeh, « Recent progress in high-entropy alloys », *Ann. Chim. Sci. Matér.*, vol. 31, nº 6, p. 633-648, déc. 2006, doi: 10.3166/acsm.31.633-648.
[21] S. Ranganathan, « Alloyed pleasures: Multimetallic cocktails », *Curr. Sci.*, vol. 85, nº 10, 2003.
[22] V. Kouznetsov, K. Macák, J. M. Schneider, U. Helmersson, et I. Petrov, « A novel pulsed magnetron sputter technique utilizing very high target power densities », *Surf. Coat. Technol.*, vol. 122, nº 2-3, p. 290-293, déc. 1999, doi: 10.1016/S0257-8972(99)00292-3.
[23] R. Herschberg *et al.*, « From Pairwise Comparisons of Complex Behavior to an Overall Performance Rank: A Novel Alloy Design Strategy », *Metals*, vol. 14, nº 12, p. 1412, déc. 2024, doi: 10.3390/met14121412.
[24] X. Wang, Z. Liu, K. Cheng, Y. Shen, et J. Li, « Corrosion characteristics of Inconel 625 cladding layer and NiCrMoAl cladding layer in molten NaCl-KCl and NaCl-KCl-Na2SO4 », *Corros. Sci.*, vol. 221, p. 111308, août 2023, doi: 10.1016/j.corsci.2023.111308.
[25] J.-C. Brachet *et al.*, « Gaines de combustible nucléaire, procédé de fabrication et utilisations contre l'oxydation/hydruration », EP3195322B1, 26 juillet 2017
[26] J. Krejčí *et al.*, « Development and testing of multicomponent fuel cladding with enhanced accidental performance », *Nucl. Eng. Technol.*, vol. 52, nº 3, p. 597-609, mars 2020, doi: 10.1016/j.net.2019.08.015.
[27] H. Okamoto, « Supplemental Literature Review of Binary Phase Diagrams: B-Fe, Cr-Zr, Fe-Np, Fe-W, Fe-Zn, Ge-Ni, La-Sn, La-Ti, La-Zr, Li-Sn, Mn-S, and Nb-Re », *J. Phase Equilibria Diffus.*, vol. 37, nº 5, p. 621-634, oct. 2016, doi: 10.1007/s11669-016-0465-z.
[28] B. Wang, Q. Wang, N. Lu, X. Liang, et B. Shen, « Enhanced high-temperature strength of HfNbTaTiZrV refractory high-entropy alloy via Al2O3 reinforcement », *J. Mater. Sci. Technol.*, vol. 123, p. 191-200, oct. 2022, doi: 10.1016/j.jmst.2022.01.025.
[29] D. B. Miracle, M.-H. Tsai, O. N. Senkov, V. Soni, et R. Banerjee, « Refractory high entropy superalloys (RSAs) », *Scr. Mater.*, vol. 187, p. 445-452, oct. 2020, doi: 10.1016/j.scriptamat.2020.06.048.
[30] R. Isayev et P. Dzhumaev, « Interaction of a diffusion barrier from the refractory metals with a zirconium alloy and a chrome coating of an accident tolerant fuel », *Nucl. Eng. Des.*, vol. 407, p. 112307, juin 2023, doi: 10.1016/j.nucengdes.2023.112307.
[31] C. Lee *et al.*, « An experimentally driven high-throughput approach to design refractory high-entropy alloys », *Mater. Des.*, vol. 223, p. 111259, nov. 2022, doi: 10.1016/j.matdes.2022.111259.
[32] L. L. Snead, D. T. Hoelzer, M. Rieth, et A. A. N. Nemith, « Refractory Alloys: Vanadium, Niobium, Molybdenum, Tungsten », in *Structural Alloys for Nuclear Energy Applications*, Elsevier, 2019, p. 585-640. doi: 10.1016/B978-0-12-397046-6.00013-7.
[33] Y. K. Cho, S. Lee, Y. Choi, H. W. Yoon, C. Lee, et K. Lim, « Diffusion barrier performance of the tungsten intermediate layer between the chromium coating and the zirconium alloy on the accident-tolerant fuel (ATF) cladding in the eutectic temperatures », *Surf. Coat. Technol.*, vol. 537, p. 133825, oct. 2026, doi: 10.1016/j.surfcoat.2026.133825.
[34] Y. Zhong *et al.*, « Study on Nb interlayer as Cr Zr diffusion barrier layer at high-temperature », *Int. J. Refract. Met. Hard Mater.*, vol. 126, p. 106956, janv. 2025, doi: 10.1016/j.ijrmhm.2024.106956.
[35] F. Pedregosa *et al.*, « Scikit-learn: Machine Learning in Python », 2012, doi: 10.48550/ARXIV.1201.0490.
[36] L. Breiman, « Random Forests », *Mach. Learn.*, vol. 45, nº 1, p. 5-32, oct. 2001, doi: 10.1023/A:1010933404324.
[37] T. Chen et C. Guestrin, « XGBoost: A Scalable Tree Boosting System », in *Proceedings of the 22nd ACM SIGKDD International Conference on Knowledge Discovery and Data Mining*, San Francisco California USA: ACM, août 2016, p. 785-794. doi: 10.1145/2939672.2939785.
[38] T. Akiba, S. Sano, T. Yanase, T. Ohta, et M. Koyama, « Optuna: A Next-generation Hyperparameter Optimization Framework », 2019, *arXiv*. doi: 10.48550/ARXIV.1907.10902.

[39] S. Bates, T. Hastie, et R. Tibshirani, « Cross-Validation: What Does It Estimate and How Well Does It Do It? », *J. Am. Stat. Assoc.*, vol. 119, n° 546, p. 1434-1445, avr. 2024, doi: 10.1080/01621459.2023.2197686.
[40] D. J. C. MacKay, « Bayesian Interpolation », *Neural Comput.*, vol. 4, n° 3, p. 415-447, mai 1992, doi: 10.1162/neco.1992.4.3.415.
[41] M. Tipping, « Sparse Bayesian Learning and the Relevance Vector Machine », *J. Mach. Learn. Res.*, n° 1, p. 211-244, 2001.
[42] H. Hajri, J.-P. Poli, et L. Boudet, « Towards Monotonous Functions Approximation from Few Data With Gradual Generalized Modus Ponens: Application to Materials Science », in *2021 IEEE 33rd International Conference on Tools with Artificial Intelligence (ICTAI)*, Washington, DC, USA: IEEE, nov. 2021, p. 796-800. doi: 10.1109/ICTAI52525.2021.00127.
[43] O. Rousselle, J.-P. Poli, et N. B. Abdallah, « Towards an Interpretable Fuzzy Approach to Experimental Design », in *Information Processing and Management of Uncertainty in Knowledge-Based Systems*, vol. 1174, M.-J. Lesot, S. Vieira, M. Z. Reformat, J. P. Carvalho, F. Batista, B. Bouchon-Meunier, et R. R. Yager, Éd., in Lecture Notes in Networks and Systems, vol. 1174. , Cham: Springer Nature Switzerland, 2024, p. 219-232. doi: 10.1007/978-3-031-74003-9_18.